\documentclass{aa}
\usepackage{graphicx}
\begin{document}
 
%
   \title{VLT optical observations of V821 Ara(=GX339--4) in an 
   extended ``off'' state
\thanks{Based on observations performed at the European Southern Observatory, 
 Paranal, Chile}}

   \author{T. Shahbaz \inst{1}
          \and R. Fender\inst{2}
          \and P.A. Charles\inst{3}
          }

   \offprints{T. Shahbaz}

   \institute{Instituto de Astrof\'\i{}sica de Canarias E-38200 La Laguna, 
              Tenerife, Spain\\
              email: tsh@ll.iac.es   
          \and
            Astronomical Institute ``Anton Pannekoek'', University of
            Amsterdam and Center for High Energy Astrophysics, 
	    Kruislaan, 403, 1098 SJ, Amsterdam, The Netherlands \\
	    email: rpf@@astro.uva.nl
	  \and
           Department of Physics \& Astronomy, University of Southampton,
           Southampton, SO17 1BJ, United Kingdom \\
	   email: pac@astro.soton.ac.uk
             }

   \date{Received ; accepted }

\abstract{
We report on low-resolution spectroscopy of  GX339--4 during 
 its current, extended X-ray
`off' state in May 2000 (r=20.1) obtained with the VLT Focal Reducer/low dispersion
Spectrograph (FORS1). Although we do not positively detect the secondary star
in GX339--4 we  place an upper limit of 30 percent on the contribution of a
``normal'' K-type secondary star spectrum to the observed flux. Using this limit for
the observed magnitude of the secondary star, we find a lower limit for the
distance of GX339--4 to be 5.6~kpc.
          }
   \maketitle
%

\section{Introduction}

The optical counterpart of GX339-4 was identified by Doxsey et al. (1979) as a
V$\sim$18 blue star. Subsequent observations showed that it exhibited a wide
range of variability depending on its X-ray state; from  V=15.4 to 20.2 (Motch
et al. 1985; Corbet et al. 1987) when it is in the X-ray `low'  and `off'
states,  while V=16-18 (Motch et al. 1985) when it is in the X-ray `high'
state.   Simultaneous optical and soft X-ray (3-6 keV) observations 
showed a remarkable anti-correlation during a
transition from an X-ray `low' to `high' state (Motch et al. 1985), the cause
of which was unknown. However, Ilovaisky et al. (1986) showed that there are
times when the optical and X-ray fluxes are correlated.
Callanan et al. (1992) reported a
possible orbital period of 14.8 h from optical photometry.

GX339--4 is of great interest in the class of black hole X-ray binaries, 
because its black hole candidacy is established by its Cyg X--1-like X-ray
variability and multiple states, yet it is the only low-mass X-ray binary
member in its  class to be ``steady'' 
(apart from the occasional ``off'' state it is usually X-ray active)  
as opposed to  transient (i.e. systems 
which undergo episodic X-ray outbursts and which usually last for 
several months and then are X-ray quiet for many years; see e.g. Charles 1998).
This may be related to the mass of the compact object but, at present,
there is no dynamical mass estimate available  (which would establish its
black-hole nature),  since there has been no spectroscopic detection of the
mass-losing star.  This is very difficult during X-ray ``on'' states due to
the brightness of the X-ray irradiated disc, but GX339--4 occasionally enters
an extended ``off'' state when the disc contribution is greatly reduced. In
this letter we report on VLT medium resolution spectroscopy of GX339--4 taken
during its current ``off'' state in order to search  for the spectral
signature  of the mass-losing star.

\begin{table}
\caption{Log of VLT observations}
\begin{center}
\begin{tabular}{lccccc}
Object    &  Date       & UT  & Exposure    &  Seeing     & Comments \\
          &  (2000)     & start      & time (s) & ($\arcsec$) & \\
 &  &  &  &  & \\
HD157423  &  05/11 & 08:57    & 2x3    & 0.63 & K1III  \\
HD155111  &  05/11 & 08:40    & 2x3    & 0.68 & K5III \\
HR5265    &  08/23 & 23:07    & 1x2    & 0.00 & K3III \\
HR5178    &  08/22 & 08:57    & 1x5    & 0.00 & K5III \\
HR6169    &  08/23 & 08:57    & 1x0.6  & 0.00 & K7III \\
V821 Ara  &  06/04 & 06:52    & 1200   & 0.74 & GX339--4 \\
V821 Ara  &  06/04 & 07:16    & 1200   & 0.69 & GX339--4 \\
\end{tabular}
\end{center}
\end{table}

\section{Observations and Data reduction}

\begin{figure} 
\begin{center}
\begin{picture}(10,300) 
\put(0,0){\includegraphics{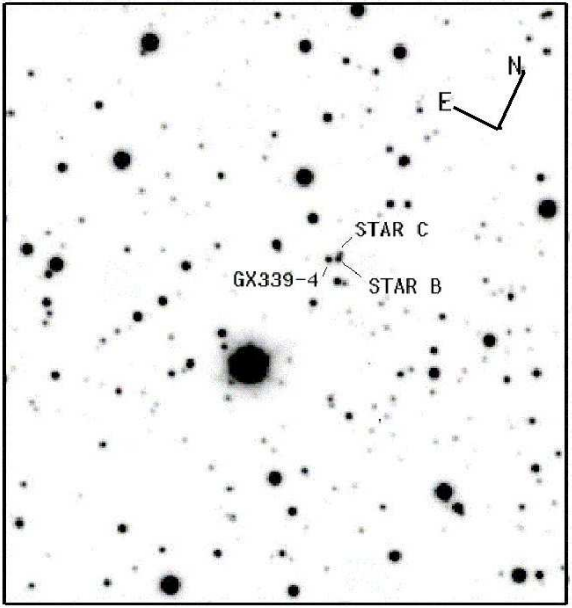}} 
\end{picture}
\caption[{\bf Figure 1}]{
R-band image of GX339--4 taken with the VLT on 4th June 2000. 
The exposure time was 30 secs and the field of view is $2\times$2 arcmins. 
For GX339--4 r=20.1$\pm$0.1. North and East are marked.}
\end{center}
\end{figure}

\subsection{Imaging}

We obtained r-band Gunn images of GX339--4  prior to the spectroscopic
observations for acquisition purposes. The integration time was 30 secs and the
images were corrected for the bias level and flat-fielded in the standard way.
We performed optimal aperture photometry (Naylor 1998) on GX339--4,  which is
clearly resolved (see Figure 1), and several nearby comparison stars. The
seeing during these observations was $\sim$0.7 arcsecs.  We calibrated the
data using photometric standard stars taken on the same night as part of the
ESO calibration programme. We find r=20.1$\pm$0.1 for GX339--4.

\begin{figure} 
\begin{center}
\begin{picture}(100,200) 
\put(0,0){\includegraphics{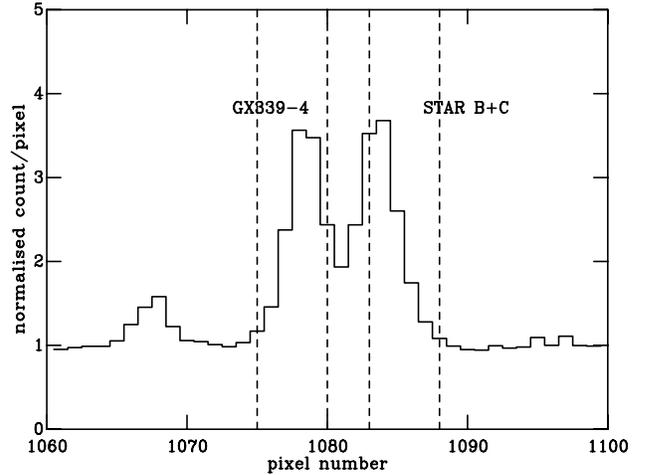}} 
\end{picture}
\caption[{\bf Figure 2}]{
A cut along the slit showing the spatial profile of GX339--4 and the
blend of stars B+C. Only rows 1075 to 1080 which are clear from contamination 
were used in the extraction of the spectrum.}
\end{center}
\end{figure} 

\subsection{Spectroscopy}
  
Spectroscopic observations of  GX339--4 were obtained using ESO's Very Large
Telescope (VLT) Unit Telescope 1 equipped with the Focal Reducer Low
Dispersion Spectrograph FORS1 (see Table 1 for a log of the observations). The
data were obtained in service mode and were taken with the 600R grism, a 0.7
arcsec slit and the standard resolution collimator, resulting in a spectral
resolution of 3.4 \AA\ (FWHM). The dispersion was 1.08~\AA\ per pixel and we
obtained a wavelength coverage from 5225--7417\AA.  Some template stars were
obtained during a  different observing run in August 2000 using the same
setup.

Acquisition images (see section 2.1) of the GX339--4 field (see Figure 1), 
clearly revealed the star 1.06
arsecs  North-West of GX339--4 in the chart published by 
Callanan et al. (1992). However, under the excellent observing conditions, 
our new image reveals
that this object can actually be resolved into two stars. These stars were close enough
to pose a potential  problem in the spectroscopic data analysis, and  
so the slit was rotated
to a position angle of 112.8 degrees East of North, to avoid them contaminating
the GX339--4 spectrum. 

The data reduction and analysis was performed using the Starlink {\sc figaro}
package, the {\sc pamela} routines of K.\,Horne and the {\sc molly} package of
T.\,R.\ Marsh. Removal of the individual bias signal was achieved through
subtraction of a median bias frame. Small scale pixel-to-pixel sensitivity
variations were removed with a flat-field frame prepared from observations of a
tungsten lamp.  Calibration of the wavelength scale was achieved using 4th
order polynomial fits. The stability of the final calibration was verified with
the OH sky lines at 6300.3\AA\ and 6363.8\AA\ whose position was accurate to
within 0.6 \AA.

One-dimensional spectra were extracted using the optimal-extraction algorithm
of Horne (1986). However, to minimize contamination from the blend of stars B
and C,  we extracted only a few rows of the CCD; as shown in Figure 2, these
were rows 1075 to 1080. The signal-to-noise ratio of the final  extracted
GX339--4 spectrum was $\sim$50 in the continuum.

\begin{figure} 
\begin{center}
\begin{picture}(100,250) 
\put(0,0){\includegraphics{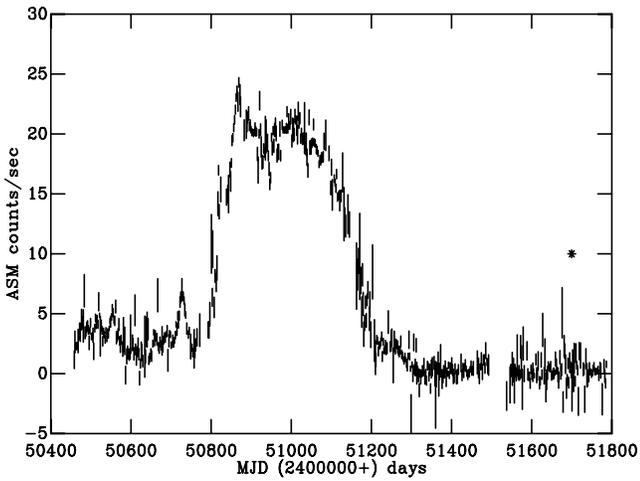}} 
\end{picture}
\caption[{\bf Figure 3}]{
RXTE ASM soft X-ray (2-10 keV) light curve of GX339--4 spanning 
the period 1 January 1997 -- 1 September 2000 (1 Crab $\simeq$ 75 cts/s). 
The star marks the time of our VLT spectroscopic observations.}
\end{center} 
\end{figure}

\begin{figure*} 
\begin{center}
\begin{picture}(100,300) 
\put(0,0){\includegraphics{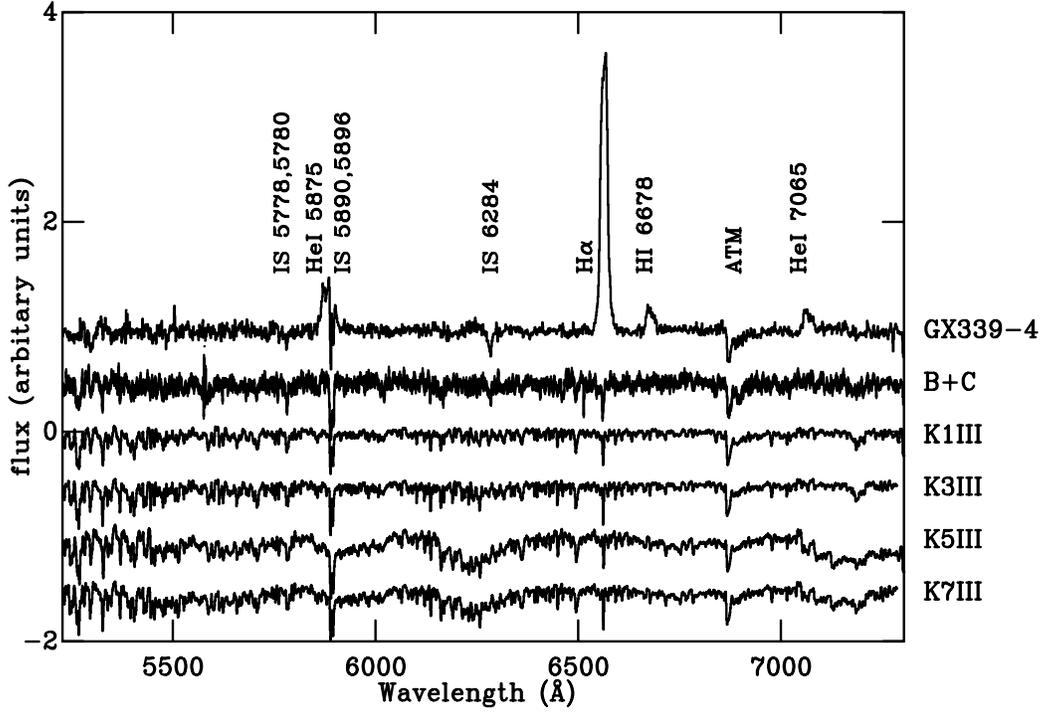}} 
\end{picture}
\caption[{\bf Figure 4}]{
From top to bottom: Variance-weighted average spectrum of GX339--4, the
blend of stars B+C; K1III, K3III, K5III and K7III 
template stars. The
spectra have been normalized and shifted vertically for clarity. IS and ATM 
indicate interstellar and atmospheric features respectively.}
\end{center}
\end{figure*}

\section{The X-ray state}
 
The RXTE All-Sky Monitor (ASM) has been continuously observing bright X-ray
sources in the $\sim$2 to 10 keV range since 1996 February 20.  The X-ray
observations (one-day averages)  of GX339--4 presented herein are extracted
from the ``quick-look results'' public archive provided on the World Wide Web
by the ASM/RXTE team (Figure 3).  During the time of our VLT observations,
GX339--4  was not detected by the ASM. SAX observations during  March 2000
revealed the source to be 3 times fainter than in August 1999 (E. Kuulkers, 
private communication)

\section{The emission lines}

The H$\alpha$, HeI~$\lambda$5876, HeI~$\lambda$6678, HeI~$lambda$7065
emission lines are clearly visible in the weighted-average spectrum of
GX339--4. The H$\alpha$ line is asymmetric, with its peak skewed
towards the red with a round top (see Figs 4 and 5). It is difficult to
say convincingly whether the profile is double-peaked, given the low
resolution of the data. However, if we assume that it is, then a
double Gaussian fit gives a peak-to-peak separation of
448$\pm$10~km~s$^{-1}$(=9.8 \AA) which is consistent with the values
of 7.4\AA\ and 8.0\AA\ obtained by Soria et al. (1999) and 
Smith, Filippenko \& Leonard  (1999) respectively when 
GX339--4 was in an ``active'' state.  The
observed H$\alpha$ equivalent width of $\sim 56$\AA\ (see Table 2) is a
factor of $\sim$8 larger than when the source is in
its high state, and is similar to that observed in the soft X-ray
transients when they are in their quiescent state
(40--60\AA). The HeI~$\lambda$5876 line appears to be double-peaked,
although it is contaminated by the NaI interstellar absorption
doublet (see Fig 5). Fitting a double-gaussian to the HeI line
gives a peak-to-peak separation of $\sim$600~km~s$^{-1}$, similar to
that observed in the HeII~$\lambda$4686 emission line when GX339--4 is in
its low-hard state (Wu et al. 2001). 

\begin{table}
\caption{Emission line equivalent widths for GX339--4}
\begin{center}
\begin{tabular}{lc}
Line            &  EW (\AA)     \\
                &                 \\
HeI~$\lambda$7065  &  -5.2$\pm$ 0.2  \\
HeI~$\lambda$6678  &  -4.8$\pm$ 0.2  \\
H$\alpha$~$\lambda$6563  & -55.9$\pm$ 0.2  \\
\end{tabular}   
\end{center}     
\end{table}      

Wu et al. (2001) have recently suggested that the H$\alpha$ profiles are
different in the high-soft and low-hard states; the high-soft state being
characterized by a double-peaked profile which arises from the irradiated
accretion disc, whereas in the low-hard state the single-peaked profiles arise
from an outflow.  However, it should be noted that if there is a bright spot
in GX339--4 then H$\alpha$ emission from this would fill-in the double-peaked
profile arising from the accretion disc, resulting in a single-peaked
profile.  A bright spot is a common feature in most of the X-ray transients 
(e.g. A0620--00; Marsh, Robinson \& Wood 1993). 
Phase-resolved high spectral resolution data will be necessary to investigate
this further.

\section{Upper limit to the secondary star contribution}

The average density of a star that fills its Roche lobe is determined solely
by its orbital period.  Assuming that the  orbital period of GX339--4 is 14.83
hr (Callanan et al. 1992) and that the secondary star does fill its Roche
lobe, then the mean density of the star is $\rho$=0.5 g~cm$^{-3}$.  A K1 main
sequence star would have $\rho$=1.9 g~cm$^{-3}$, implying that the radius of
the secondary must be a factor of 1.6 greater than  that of a main
sequence star in order for it to fill its Roche lobe.  It is therefore most
likely to be evolved, similar to the secondary star in Cen X--4 (Shahbaz,
Naylor \& Charles 1993). 

The GX339--4 spectrum does not show any signs of obvious absorption features 
from the secondary star (see Figure 4). However, it is still possible to
determine an $\it upper$ limit to the contribution of such a star (Dhillon et
al. 2000).  This is done by subtracting a constant times the normalized
template spectrum from the normalized GX339--4 spectrum, until spectral 
absorption features  from the template star appear in emission in the GX339-4
spectrum.  The value of the constant $\it f$ at this point represents an upper
limit to the fractional contribution of the secondary star.  The contribution
also depends on the spectral type of the template used.  As the spectral type
of the secondary star in GX339--4 is unknown, we used stars in the range
K1III--K7III and find upper limits to the secondary star 
contribution to lie in the range 20--30\%. It should be noted that giant stars
have weaker metallic  absorption lines (near H$\alpha$) compared to sub-giant
stars. This implies that our estimates for the secondary star's contribution
are upper limits. 

\begin{figure*} 
\begin{center}
\begin{picture}(150,300) 
\put(0,0){\includegraphics{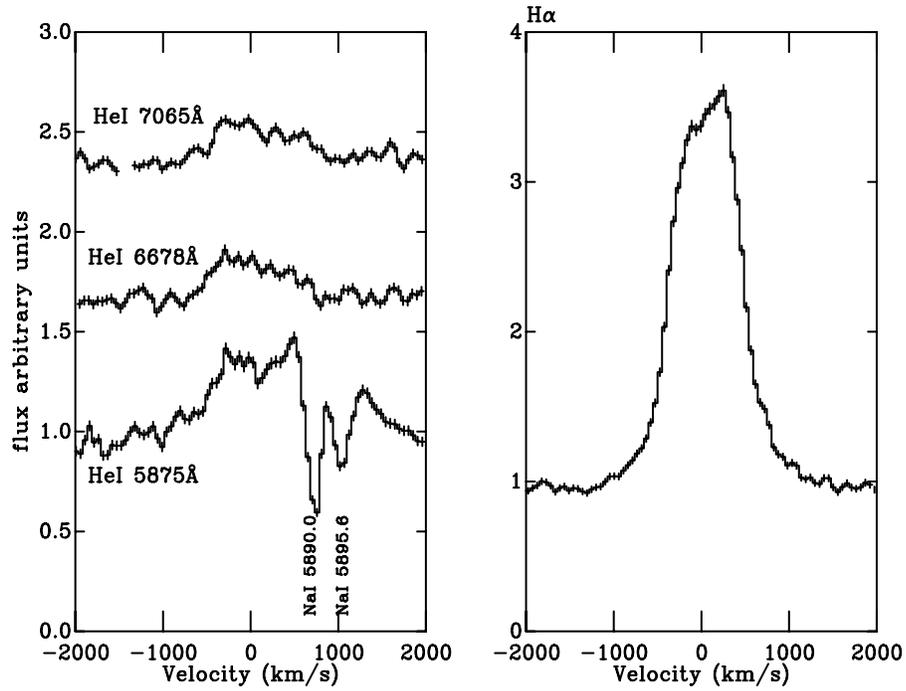}} 
\end{picture}
\caption[{\bf Figure 5}]{
Left: Close-up of the HeI 5876\AA  + NaI  interstellar
absorption doublet, 6678\AA\ and 7065\AA\ emission lines. The spectra have been
normalized and offset for clarity.
The  HeI 5876\AA  line is clearly double-peaked. Right: A close-up of
the H$\alpha$ emission line. The profile has a round top profile. 
The low resolution
of the data makes it difficult to say whether the profile is single- 
or double-peaked.} 
\end{center}
\end{figure*}

\section{Lower limit for the distance}

We can place a lower limit to the distance of GX339--4, by comparing
our observed upper limit for the secondary star's magnitude with that
expected for a Roche-lobe filling secondary star in orbit around a
black hole. Using the observed magnitude of r=20.1 (the H$\alpha$
emission line contribution to the observed flux is negligible and so
all the light is assumed to arise from the secondary star) and our
upper limit for the secondary star's contribution to the observed
light, $f<$30\%, we find that the secondary star must have a r-band
magnitude fainter than 20.4. Using the ellipsoidal model described in
Shahbaz, Naylor \& Charles (1993) we determine a lower limit to the
magnitude of the secondary star and the distance to the
source. Although there were no X-ray measurements during the time of
Callanan's optical observations, the X-ray luminosity of GX339--4 in
its ``off'' state, within a few years of their observations was
reported to be $\sim 2\times 10^{35}$~erg~s$^{-1}$ (Ilovaisky et
al. 1986).  If we assume that the X-ray luminosity of GX339--4 in its
``off'' state was similar, then the optical light curve of a system
with this X-ray luminosity would be dominated by X-ray heating and
thus would appear single-humped and would be modulated on the orbital
period. Hence, the interpretation of Callanan et al.  (1992) that the
modulation they observe is the orbital period would seem correct.  If
the orbital period is 14.8 hrs (Callanan et al. 1992) and assuming a
binary mass ratio of 10 (black hole mass/secondary star mass), a black
hole of 10 M$_{\odot}$ (median mass observed; Miller, Shahbaz \& Nolan
1998), an inclination of 15 degrees (Wu et al. 2001) and a mean
secondary star temperature of 4300K [c.f. the secondary star in Cen
X--4 (Chevalier et al.  1989) which has a similar orbital period to
GX339--4] and a colour excess of E$_{B-V}$=1.2 (Zdziarski et al. 1998)
we find a lower limit to the distance of 5.6~kpc. This value is
consistent with crude distance estimates determined from the systemic
velocity of GX339--4; 4$\pm$1~kpc (Zdziarski 1998).

The lack of absorption line features in our GX339--4 spectrum is
puzzling. One possibility, that cannot be ruled out is that the
secondary star has a much earlier spectral type. We can use the
average scale height of black hole X-ray binaries and the Galactic
latitude to estimate the distance to GX339--4. Using a scale height of
$\sim$500pc (White \& van Paradijs 1996) and b=-4.3 degrees, the
distance is $\sim$7 kpc. At this distance, and given our observed
magnitude, the secondary star would require a spectral type later than F8
(i.e.  $<$6200~K). Note that the optical spectrum of such a star near
H$\alpha$ would appear featureless, i.e. one would not expect to see
strong absorption lines.


\begin{acknowledgements}

TS was supported by a EC Marie Curie Fellowship HP-MF-CT-199900297.

\end{acknowledgements}

\end{document}